\documentclass[fleqn]{article}
\begin{document}
\thispagestyle{empty}
 \begin{center}
\begin{large}
 {\bf {  A NEW CURRENT REGURALIZATION OF THIRRING MODEL}}

\end{large}

\vspace{1cm}

Hidenori TAKAHASHI\footnote{e-mail: htaka@phys.ge.cst.nihon-u.ac.jp}

Laboratory of Physics, College of Science and Technology

Nihon University, Funabashi Chiba 274-8501, Japan 

\vspace{5mm}
and
\vspace{5mm}

Akihiro OGURA\footnote{e-mail: ogu@mascat.nihon-u.ac.jp}

Laboratory of Physics, School of Dentistry at Matsudo 

Nihon University, Matsudo, Chiba 271-8587, Japan

\vspace{2cm}

{\Large ABSTRACT}

\end{center}

We study an ambiguity of the current regularization in the Thirring model.
We find a new current definition which enables to make a comprehensive treatment
of the current. Our formulation is simpler than Klaiber's formulation.
We compare our result with other formulations and find
 a very good agreement with their result. We also obtain the Schwinger term and
the general formula for any current regularization.

\section{Introduction}
The Thirring model has been investigated by many people. 
It is well-known that the Thirring model is an exactly solvable quantum field theory model
in (1+1) dimensions \cite{thi58}.
An extensive investigation of the model was given by Hagen \cite{hagen1967}
and Klaiber \cite{kla67}.
Hagen introduced an external field and gave the general solution of the Thirring model.
Klaiber analyzed the Thirring model and found the operator solutions which are
expressed in terms of a free massless Dirac field.
He constructed the solution to fulfill the positive definiteness.
On the other hand, Nakanishi expressed the solution in terms of 
the free massless bosonic field \cite{nak77}.
He asserted that {\it all Heisenberg operators
 should be expressed in terms of asymptotic fields} from the standpoint of
 the general principle of quantum field theory.  In the present paper, 
we use the bosonic expression ({\it bosonization}) \cite{stone94,ogu94}. 

One of the methods for solving the quantum field theory is to determine the Operator Product 
Expansion (OPE) \cite{wilson69,kad69}. In this formalism, the short distance behavior 
for products of the two local fields is important. For some case, we can determine the 
OPE exactly, e.g. (1+1)-dimensional Conformal Field Theory \cite{bel_pol_zam84}.
Concerning the operator products of the quantum field,
there are difficulties with respect to the current regularization. 
In most cases, the current is defined by the limiting procedures as 
$\bar\psi(x+\epsilon)\gamma_\mu\psi(x)$. In the Thirring model, there are 
several definitions of the current. 
For example, the Schwinger current \cite{sch59} is defined by limiting from
spacelike direction only. The Johnson current \cite{joh61}
is defined by limiting from not only spacelike direction but
 also timelike direction symmetrically as
\begin{equation}
j^\mu (x) = \frac{1}{2} \biggl[j^\mu(x;\, \epsilon)
          + j^\mu(x;\, \widetilde\epsilon) 
         \biggr]_{\epsilon, \widetilde\epsilon \rightarrow 0},
\end{equation}
where $\epsilon$ and $\widetilde\epsilon$ are a timelike and spacelike
vectors, respectively.
Both current definitions are consistent with the solution of the Thirring model.
However, the coupling constant is affected by the current definition.
Therefore, in the Thirring model, the coupling constant is determined only when 
we define the current regularization \cite{kla67}. It is also noted that
these coupling constants are not independent. The coupling constant of 
the Schwinger definition $g_{\rm S}$ is given by
\begin{equation}
g_{\rm S} = \frac{g_{\rm J}}{1 - g_{\rm J}/2 \pi},
\end{equation}
where $g_{\rm J}$ is the coupling constant of the Johnson definition.
These current ambiguities also appear in the massive Thirring model, which
we do not understand yet \cite{fkt2000}.

On the other hand, Dell'Antonio, Frishman and Zwanziger \cite{del_fri_zwa} 
analyzed the Thirring model without looking into the structure of the current.
They extend the Johnson result \cite{joh61}.
They start with defining the commutation relations of the current, 
{\it current algebra formulation} \cite{wilson69,kad69}.
There are three parameters and we have two relations among them.
Therefore, we can construct the current algebra from the lagrangian and 
the suitable definition of the current which has one parameter.

In this paper, we present a new current regularization of the Thirring model.
We introduce one parameter in the definition. Our formulation is simpler
than Klaiber's one and the new current definition 
is consistent with other formulations.
The Thirring current and field can be written in terms of the free massless bosonic field.
Therefore, we can analyze the model exactly.

In this paper, we employ the following notation:
\begin{equation}
x^\pm = x^0 \pm x^1,\; x_\pm = x^\mp/2, \;
\partial_\pm = (\partial_0 \pm \partial_1)/2, \;
\partial^\pm = 2 \partial_\mp
\end{equation}
and gamma matrices are
\begin{equation}
\gamma^0 = \pmatrix{   0 &  1  \cr 
                       1 &  0   }, \quad
\gamma^1 = \pmatrix{   0 & -1  \cr 
                       1 &  0  \cr }, \quad
\gamma^5 = \gamma^0 \gamma^1 = \pmatrix{   1 &  0  \cr 
                                           0 & -1  \cr }.
\end{equation}
The anti-symmetric tensor $\epsilon_{\mu \nu}$ is taken to be
 $\epsilon_{10} = \epsilon^{01}=-1$.

\section{Thirring model}
The Thirring model is (1+1) dimensional field theory with the current-current 
interaction. 
The lagrangian of the Thirring model is given by
\begin{equation}
{\cal L}= \bar\psi{\rm i} \partial_\mu \gamma^\mu \psi 
 -  \frac{g}{2} j_ \mu j^\mu, \quad \quad 
j_\mu = \bar\psi\gamma_\mu\psi,
\end{equation}
where $g$ is a coupling constant. Then, the equations of motion become
\begin{equation}
     {\rm i} \partial_+ \psi_{\rm R} = g j_{\rm L} \psi_{\rm R}, \quad
     {\rm i} \partial_- \psi_{\rm L} = g j_{\rm R} \psi_{\rm L},
\label{thir_eq200506}
\end{equation}
where
\begin{equation}
\psi^{\rm T} = (\psi_{\rm R},\;\psi_{\rm L}), \quad
j_{\rm R} \equiv \frac{1}{2} \left(j^0 + j^1 \right), \quad 
j_{\rm L} \equiv \frac{1}{2} \left(j^0 - j^1 \right).
\end{equation}
From eq.(\ref{thir_eq200506}), the current $j^\mu$ and its dual (axial) current
$\widetilde{j}^\mu = \epsilon^{\mu \nu} j_\nu$ is conserved,
\begin{equation}
\partial_\mu j^\mu =0, \quad \partial_\mu \widetilde{j}^\mu = 0.
\end{equation}
The Thirring model is exactly solvable.
Thirring \cite{thi58} constructed the eigenstates while
Klaiber \cite{kla67} found the operator solution.
On the other hand, Nakanishi \cite{nak77} described the quantum operator solution of 
eq.(\ref{thir_eq200506}) in terms of the free massless bosonic field 
$\varphi= \varphi_{\rm R}(x^-) + \varphi_{\rm L}(x^+)$ as
\begin{equation}
\psi(x) = \frac{Z}{\sqrt{2 \pi}}\pmatrix{ \mbox{\Large :} 
e^{{\rm i}\, s \varphi_{\rm R} - {\rm i}\,\bar{s} \varphi_{\rm L}} \mbox{\Large :} \cr
\mbox{\Large :} e^{{\rm i}\, \bar{s} \varphi_{\rm R} - {\rm i}\,s \varphi_{\rm L}}
 \mbox{\Large :}},
\label{thir_op200509}
\end{equation}
where $s, \bar{s}$ are constant parameters and $Z$ is a normalization factor.
The free bosonic field satisfies
\begin{equation}
\partial_\mu \partial^\mu \varphi =0
\end{equation}
and we can regularize \cite{abd2_rot} as
\begin{equation}
\left[\varphi_{\rm R}^\downarrow (x^{-}), 
        \varphi_{\rm R}^\uparrow (y^{-})\right] 
= - \frac{1}{4 \pi} \ln  {\rm i}(x^-- y^--{\rm i} 0),
\end{equation}
\begin{equation}
\left[\varphi_{\rm L}^\downarrow (x^{+}), 
        \varphi_{\rm L}^\uparrow (y^{+})\right] 
= - \frac{1}{4 \pi} \ln  {\rm i}(x^+- y^+-{\rm i} 0),
\end{equation}
where $\varphi_{\rm R, L}^\downarrow$ and $\varphi_{\rm R, L}^\uparrow$ are
the positive and the negative frequency part respectively. 
Therefore, we have the Operator Product Expansion (OPE) of the Thirring operator, 
\begin{eqnarray}
\psi_{\rm R}(x) \psi_{\rm R}(y) &=& \frac{|Z|^2}{2 \pi}
                    {\rm i}^{\frac{s^2+\bar{s}^2}{4 \pi}}
                    (x^- - y^- - {\rm i} 0)^{s^2/4\pi}
                    (x^+ - y^+ - {\rm i} 0)^{\bar{s}^2/4\pi} \nonumber\\
& & \quad \quad \times \mbox{\Large :} 
  e^{{\rm i}\, s \varphi_{\rm R}(x) - {\rm i}\,\bar{s} \varphi_{\rm L}(x)
                + {\rm i}\, s \varphi_{\rm R}(y) - {\rm i}\,\bar{s} \varphi_{\rm L}(y)}
 \mbox{\Large :}, \\
\psi_{\rm R}^\dagger(y) \psi_{\rm R}(x) &=& \frac{|Z|^2}{2 \pi}
                    {\rm i}^{-\frac{s^2+\bar{s}^2}{4 \pi}}
                    (y^- - x^- - {\rm i} 0)^{-s^2/4\pi}
                    (y^+ - x^+ - {\rm i} 0)^{-\bar{s}^2/4\pi} \nonumber\\
& & \quad \quad \times \mbox{\Large :} 
  e^{ -{\rm i}\, s \varphi_{\rm R}(y) + {\rm i}\,\bar{s} \varphi_{\rm L}(y)
                + {\rm i}\, s \varphi_{\rm R}(x) - {\rm i}\,\bar{s} \varphi_{\rm L}(x)}
 \mbox{\Large :} 
\end{eqnarray}
and so on. We have the similar relation for $\psi_{\rm L}$ if $\bar s \leftrightarrow s$.
For the massless Dirac field case ($g=0$), we find $s=2 \sqrt{\pi}$ and $\bar{s}=0$.

To solve the model, we must determine the parameters $s, \bar{s}$.
The first condition of $s$ and $\bar{s}$ is given by the aniti-commutativity of
$\psi$ and we have
\begin{equation}
\frac{s^2 - \bar{s}^2}{4 \pi} = 1.
\end{equation}

\section{Current regularization}
Next, we insert the operator solution into the field equation 
eq.(\ref{thir_eq200506}).
To do this, we propose the following current definition,
\begin{eqnarray}
j^\mu (x) &=& \frac{1}{2} \biggl[ \bar{\psi}(x+\varepsilon) \gamma^\mu \psi(x)
       + \bar{\psi}(x+\widetilde\varepsilon) \gamma^\mu \psi(x) \biggr] \nonumber\\
      & & \quad
       - \frac{\sigma}{2} \alpha^\mu_{\;\;\nu} \biggl[ \bar{\psi}(x+\varepsilon) \gamma^\nu \psi(x)
       - \bar{\psi}(x+\widetilde\varepsilon) \gamma^\nu \psi(x) \biggr],
\end{eqnarray}
where $\varepsilon(\widetilde\varepsilon)$ is an infinitesimal timelike (spacelike) vector and
$\sigma$ is a parameter and $\alpha^0_{\;\;0} =- \alpha^1_{\;\;1}=1$, 
$\alpha^0_{\;\;1}=\alpha^1_{\;\;0}=0$. 
Here, we get timelike vector $\varepsilon^\mu$ close to zero 
with $\varepsilon^1 \rightarrow 0$ firstly,
whereas the spacelike vector $\widetilde\varepsilon$ is done in an opposite way.
In our formulation, the current is written by
\begin{equation}
j_{\rm R} = - \frac{s - \sigma \bar{s}}{2 \pi} \partial_- \varphi_{\rm R}, \quad
j_{\rm L} = \frac{s - \sigma \bar{s}}{2 \pi} \partial_+ \varphi_{\rm L}.
\end{equation}
Therefore, the operator solution (\ref{thir_op200509}) is valid if
\begin{equation}
\bar{s} = \frac{g}{2 \pi}(s - \sigma \bar{s}).
\end{equation}
Finally, we have the equations which must be satisfied by the parameters of the solution
\begin{equation}
\frac{s^2 - \bar{s}^2}{4 \pi} = 1, \quad \bar{s} = \frac{g}{2 \pi}(s - \sigma \bar{s}).
\end{equation}
First, we consider $\sigma=0$ and $\sigma=1$ case. The second equation becomes
\begin{equation}
\bar{s} = \frac{g_{\sigma=0}}{2 \pi} s, \quad \bar{s} = \frac{g_{\sigma=1}}{2 \pi}(s - \bar{s}) .
\end{equation}
They can identify with the map,
\begin{equation}
g_{\sigma=1} = \frac{g_{\sigma=0}}{1- g_{\sigma=0}/ 2 \pi} .
\end{equation}
This is nothing but the relation between the coupling constant of the Schwinger
definition and that of the Johnson definition. This is consistent with 
Klaiber's result \cite{kla67}. 
Therefore, $\sigma=1$ corresponds to  Schwinger's current definition and
$\sigma=0$ is  Johnson's one in our formulation.

We can also calculate the commutation rules between the current and 
the spinor field $\psi$,
\begin{equation}
\biggl[\psi(x^1, t), \;j^0(y^1,t)\biggr] = 
\frac{(s - \sigma \bar s)(s + \bar s)}{4 \pi} \delta(x^1 - y^1) \psi(x)
\label{com_cur1_200512}
\end{equation}
and
\begin{equation}
\biggr[\psi(x^1, t), \;j^1(y^1, t)\biggr] = 
\frac{(s - \sigma \bar s)(s - \bar s)}{4 \pi} \delta(x^1 - y^1) \gamma^5 \psi(x).
\label{com_cur2_200512}
\end{equation}

\section{Comparison with other formulations}
Dell'Antonio, Frishman and Zwanziger \cite{del_fri_zwa} 
analyzed the Thirring model in a different way. They consider
the commutation relations of the current, the Schwinger term.
 We can identify their result with
\begin{equation}
a = \frac{(s - \sigma \bar s)(s + \bar s)}{4 \pi}, \quad
\bar a =  \frac{(s - \sigma \bar s)(s - \bar s)}{4 \pi}, \quad
c = \left(\frac{s - \sigma \bar s}{2 \pi} \right)^2,
\end{equation}
where $a$, $\bar{a}$ and $c$ are parameters in their formulation 
(Note that in \cite{del_fri_zwa}, $\epsilon_{\mu\nu}$ is defined by $\epsilon_{10}=1$).
It is easy to check the consistency condition, eq.(6.1) in \cite{del_fri_zwa}, 
\begin{equation}
a - \bar{a} = g c .
\end{equation}
$c$ is written in terms of the coupling constant $g$ as
\begin{equation}
 1/c = \pi \left[ 1 + \frac{g}{2 \pi} (\sigma - 1) \right]
           \left[ 1 + \frac{g}{2 \pi} (\sigma + 1) \right].
\label{c_g}
\end{equation}
If $\sigma =0$, it becomes eq.(6.3) in \cite{del_fri_zwa}. Therefore, our result 
perfectly agrees with Dell'Antonio et al. and the parameter of the current commutation relation is 
determined by the coupling constant and the parameter $\sigma$ 
appeared in  the current definition.

Taguchi, Tanaka and Yamamoto \cite{tag_tan_yam} consider the Thirring model 
with the Tomonaga-Schwinger equation. 
They consider the deformed hamiltonian and calculate the commutation relations 
between the current and the spinor field. 
In this case, we have eq.(\ref{com_cur1_200512}) and (\ref{com_cur2_200512}) 
in a similar way.

It is well-known that the Thirring model is $c=1$ ($c$ is the central charge)
 Conformal Field Theory (CFT) \cite{fer_gri_gat72}. Klassen and Melzer \cite{kla_mel93}
argued that the Thirring model is equivalent to the {\it fermionic} Gaussian CFT.
They show the relation between the compactification radius of the fermionic Gaussian CFT 
and the Thirring coupling constant. We give their result with $\sigma=0$ (Johnson current).
More generally, the compactification radius $R$ is written by
\begin{equation}
R = \frac{1}{\sqrt{\biggl[1+ \displaystyle\frac{g}{2 \pi}(\sigma-1)\biggr]
                   \biggl[1+ \displaystyle\frac{g}{2 \pi}(\sigma+1)\biggr]}}\biggl[
      1 \pm \frac{g}{2 \pi} \sqrt{1-\frac{4 \pi}{g} \sigma - \sigma^2} \;\biggr] .
\end{equation}

\newpage
\section{Conclusion}
We have presented the generalization of the current regularization in the Thirring model.
The definition of the current is complicated, 
but it becomes simple when it is expressed in terms of the free massless bosonic field.
The present description is consistent with known results of the Thirring model.
The present formulation is simpler than Klaiber's formulation. 
Klaiber defines the commutator of the current $j^\mu$ and the field $\psi$,
and makes the anzatz about the current while we employ
the operator solution which is given by Nakanishi \cite{nak77}.
 The solution is written in terms of the {\it free massless bosonic field}, and thus
we can easily evaluate various quantities. This is the main difference between Klaiber's
treatment and ours. Further, we obtain the general formula for arbitrary current regularization.

The short distance behavior of the Thirring model is more complicated than the free massless
Dirac field. For the Dirac field, the limiting procedures of $\varepsilon \rightarrow 0$
and $\widetilde\varepsilon \rightarrow 0$ are the same. On the other hand, they are different
for the Thirring field. This is the consequence of the fact that the Dirac
field ($\bar{s} =0$) is written in terms of the bosonic field $\varphi_{\rm R}$ and 
$\varphi_{\rm L}$ separately in contrast with the Thirring field. 

In the present description, we introduce a new parameter $\sigma$ in our current definition.
This current becomes Lorentz covariant limiting operator,
 $j^\mu \sim \partial^\mu \widetilde\varphi$. 
Here, $\widetilde\varphi$ is the dual massless field of $\varphi$\cite{abd2_rot}.
We obtain the Johnson current for $\sigma=0$ and the Schwinger current for $\sigma=1$. 
Note that, for $\sigma=1$, the current definition is given by
\begin{equation}
j^0 (x) = \bar{\psi}(x+\widetilde\varepsilon) \gamma^0 \psi(x), \quad \quad
j^1 (x) = \bar{\psi}(x+\varepsilon) \gamma^1 \psi(x).
\end{equation}
Therefore, in our formulation, $j^0$ is defined by limiting from spacelike direction
while $j^1$ is defined by limiting from timelike direction.
This is in contrast with the original Schwinger's definition \cite{sch59} which 
is defined by the spacelike separation only.
 However, we can show that $j^0$ and $j^1$ commute with each other for the Thirring case
 if we adopt the original prescription.
On the other hand, in our formulation, the Schwinger term \cite{got_ima,sch59} appear as
\begin{equation}
\left[j_0(x^1,t), \;j_1(y^1, t) \right] = {\rm i} \,c \, \delta^\prime (x^1 - y^1),
\end{equation}
where $c$ is given by eq.(\ref{c_g}).
Accordingly, the current must be defined by limiting from both spacelike and 
timelike direction in the Thirring model. Note that the parameter $\sigma$ determines
the current algebra.
Unfortunately, we do not understand the physical meaning of the parameter $\sigma$ yet.

\vspace{1cm}
\hspace{-5mm}{\bf ACKNOWLEDGMENTS}

\vspace{5mm}
We would like to thank T. Fujita and M. Hiramoto for helpful discussions and comments.

\end{document}